\documentclass{aa}
\usepackage{txfonts,graphicx,natbib}
\bibpunct{(}{)}{;}{a}{}{,}

\begin{document}

      \title{Morphological study of the nested planetary nebula Hubble 12\
\thanks{Reduced imaging and spectroscopic data are only available at the CDS via anonymous ftp to cdsarc.u-strasbg.fr (130.79.128.5) or via 
http://cdsarc.u-strasbg.fr/viz-bin/qcat?J/A+A/}}
     
   \author{Chih-Hao Hsia\inst{1,2}
   \fnmsep\thanks{Corresponding author, \email{chhsia@must.edu.mo}}
   \and Yong Zhang\inst{3}
   \fnmsep\thanks{Corresponding author, \email{zhangyong5@mail.sysu.edu.cn}}
   \and SeyedAbdolreza Sadjadi\inst{4}
   \and Wayne Chau\inst{4}
   \and Hui-Jie Han\inst{1,5}
   \and Jian-Feng Chen\inst{1}
    }

   \titlerunning{Multi-wavelength study of planetary nebula Hb 12}
    \authorrunning{C. -H.~Hsia et al.}
   
    \institute{State Key Laboratory of Lunar and Planetary Sciences, Macau University of Science and Technology, Macau, China 
    \and
       CNSA Macau Centre for Space Exploration and Science, Macau, China 
    \and
      School of Physics and Astronomy, Sun Yat-Sen University, 2 Da Xue Road, Tangjia, Zhuhai, 519000, Guangdong Province, China 
     \and
        Laboratory for Space Research, Faculty of Science, Department of Physics, The University of Hong Kong, Hong Kong, China 
        \email{abdi1374@connect.hku.hk} 
      \and
        Nantong Academy of Intelligent Sensing, Nantong 226000, China 
              }

   \date{Received 11 June 2021 / Accepted 26 August 2021}

\abstract{We present a visible-infrared imaging study of young planetary nebula (PN) Hubble 12 (Hb 12; PN G111.8-02.8) obtained with {\it Hubble Space Telescope} (HST) archival data and our own {\it Canada-France-Hawaii Telescope} (CFHT) measurements. Deep HST and CFHT observations of this nebula reveal three pairs of bipolar structures and an arc-shaped filament near the western waist of Hb 12. The existence of nested bipolar lobes together with the presence of H$_{2}$ knots suggests that these structures originated from several mass-ejection events during the pre-PN phase. To understand the intrinsic structures of Hb 12, a three-dimensional model enabling the visualisation of this PN at various orientations was constructed. The modelling results show that Hb 12 may resemble other nested hourglass nebulae, such as Hen 2-320 and M 2-9, suggesting that this type of PN may be common and the morphologies of PNs are not so diverse as is shown by their visual appearances. The infrared spectra show that this PN has a mixed chemistry. We discuss the possible material that may cause the unidentified infrared emissions. The analyses of the infrared spectra and the spectral energy distribution suggest the existence of a cool companion in the nucleus of this object. 
}

    \maketitle

\section{Introduction}

When intermediate- and low-mass stars evolve from asymptotic giant branch (AGB) stars and pre-planetary nebulae (PPNs) into planetary nebulae (PNs), the pre-existing circumstellar envelopes are predicted to be present outside of the bright nebular shells \citep{Kwok82}. Early eruptions play a principal role in recycling materials from stars in the interstellar medium (ISM). These envelopes,  known as AGB haloes, have a low surface brightness (with a typical value of $\sim$ 10$^{-3}$ times fainter than the PN main shells) and may show high extinction because of dust, which makes them difficult to detect in the visible light. High dynamic range imaging has shown that faint optical haloes around PNs are commonly present (e.g. Chu et al. 1987; Balick et al. 1992; Corradi et al. 2003). Although most main shells of PNs show some degree of asymmetry, the surrounding haloes are generally spherical. In order to understand the morphological transformation of the ejections, it would be useful to obtain deep wide-field mapping of PNs, which offers valuable information about the mass-loss processes near the end of the AGB phase.

The $\upsilon$=1-0 S(1) molecular hydrogen line at 2.122 $\mu$m (hereafter called H$_{2}$ line) is an excellent tracer of circum-nebular materials. This line has commonly been observed in PPNs \citep{Sahai98, Hrivnak06, Hrivnak08, Gledhill12} and PNs \citep{Kastner94, Kastner96, Latter95, Guerrero00, Ramos08, Ramos17, Marquez13, Marquez15, Akras20}. This has yielded significant insight into  the mass-loss history during the AGB phase and the mechanism that shapes PNs. The H$_{2}$ line is excited through the fluorescence pumped by ultraviolet (UV) radiation from the central source \citep{Black87, Gledhill12}, thermal excitation in a merged ionisation-dissociation front \citep{Bertoldi96, Henney07}, and shocks \citep{Ramos08, Marquez15}. If the excitation mechanism is dominated by shocks, this line can be used to evaluate the evolutionary status of the nebulae \citep{Davis03}. Imaging studies of bipolar PNs with H$_{2}$ emission have revealed that the strengths of H$_{2}$ lines are correlated with their bipolar morphology \citep{Kastner94, Kastner96, Guerrero00}. Bipolar PNs with broad equatorial rings are found to have stronger H$_{2}$ emission than those with a narrow waist. The cause of this correlations is still unclear. A possible explanation is that the two types of bipolar PNs are at different evolutionary stages \citep{Guerrero00}.

\begin{figure*}
\begin{center}
\includegraphics[width=0.9\textwidth]{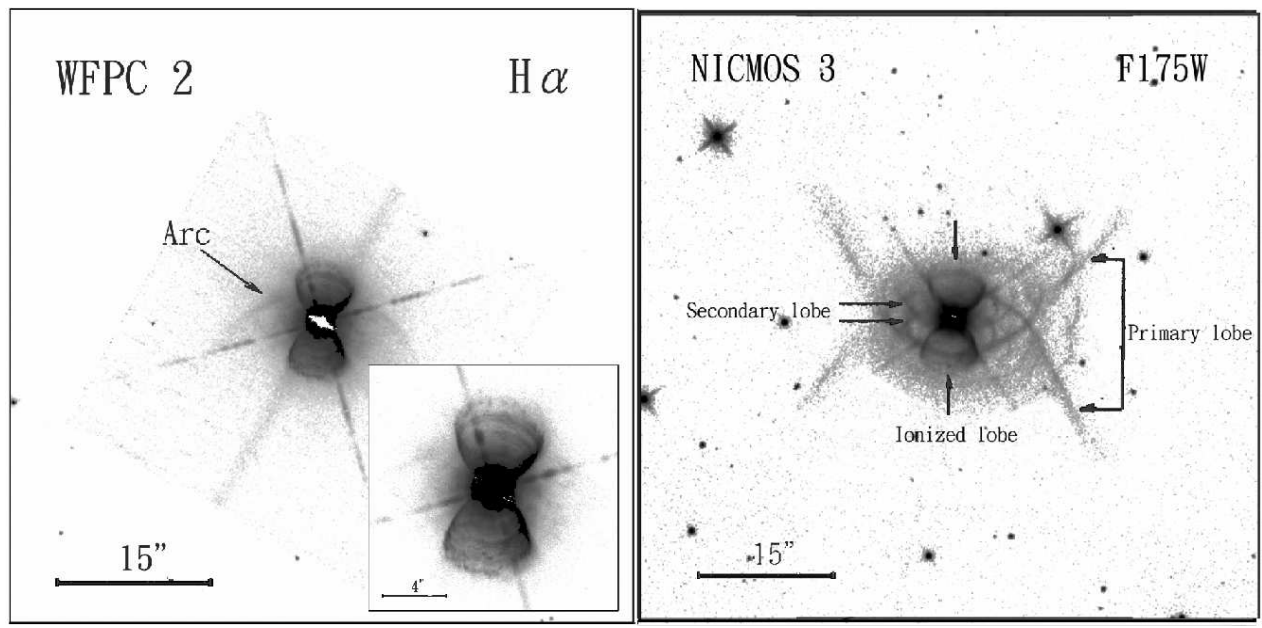}
\end{center}
\caption{Comparison between {\it HST} H$\alpha$ (left) and NIR (right) images of Hb 12. The intensities shown in the left and right panels are on logarithmic and linear grey scales, respectively. North is up and east is to left. Left: Faint arc-like filament is visible in the northern part of the nebula. This arc might be the optical counterpart of the northern part of the eye-shaped structure seen in the right panel. In the inset of this image, a series of rings aligned with bipolar lobes are visible. The central cross of this nebula is a diffraction artefact as the core is saturated. Right: Several morphological structures are labelled in the F175W image. These three bipolar lobes are labelled 'ionised lobe', 'secondary lobe', and 'primary lobe'. An eye-shaped structure is visible in the equatorial region of Hb 12.}

\label{hst}
\end{figure*}

Hubble 12 (Hb 12; PN G111.8-02.8) is classified as a young PN owing to its high surface brightness in the radio and infrared (IR) windows \citep{Aaquist90, Zhang90}. Its evolutionary age is consistent with the kinematic age of the nebula, ranging from about 300 years for the innermost structure \citep{Miranda89} to 1,120 years for entire object \citep{Vaytet09}. The precise determination of the distance for this object has been a difficult problem for a long time. Based on the estimates of different methods, the distance of the PN varies between 2.24 kpc \citep{Cahn92} and 14.25 kpc \citep{Stanghellini10}. From optical spectroscopic observations, \citet{Hyung96} found that a zone with a high electron density of $\sim$ 5$\times$10$^{5}$ cm$^{-3}$ and corresponding electron temperature of 13,600 K exists in this compact nebula. Based on a photoionisation model, \citet{Hyung96} derived that the effective temperature and luminosity of the central source are estimated to be about 35,000 K and 1,200 L$_{\sun}$. The optical appearance of this nebula is similar to that of other bipolar hourglass PNs such as MyCn 18, Mz 3, M 2-9, and Hen 2-104, all of which are known to have high-speed collimated outflows with an expanding velocity of about few hundred km~s$^{-1}$ \citep{Balick89, Bryce97, O'Connor00, Solf00, Corradi01, Guerrero04}. These high-velocity outflows have been suggested to originate from binary interactions \citep{White92, Vaytet09, Lykou11, Castro12, Misz18}.

\citet{Hora96} presented the H$_{2}$ images of Hb 12, which exhibit 
 a tight and confused structure closed to the equatorial region.
The excitation has been attributed to fluorescence \citep{Dine88, Rams93}. Moreover, a shocked [\ion{Fe}{II}] line was detected in the bipolar structures near the compact core region \citep{Welch99}, suggesting the existence of a high-speed stellar wind emanating from the central source. These observations indicate that this object is located at a short transition stage between young bipolar PNs and the evolved butterfly PNs.

To search for the possible link between the shaping of bipolar nebula Hb 12 and its central source, we study visual and near-infrared (NIR) images with high-angular-resolution of this PN, obtained from the {\it Hubble Space Telescope} (HST) and {\it Canada-France-Hawaii Telescope}\footnote{Based on observations obtained with WIRCam, a joint project of CFHT, Taiwan, Korea, Canada, France, at the Canada-France-Hawaii Telescope (CFHT) which is operated by the National Research Council (NRC) of Canada, the Institut National des Sciences de l'Univers of the Centre National de la Recherche Scientifique of France, and the University of Hawaii} (CFHT), respectively, and the IR spectra. The appearance of this nebula reveals unusual nested bipolar shells and surrounding structures. The observation and corresponding data reduction are introduced in Sect. 2. In Sect. 3 we present our spectroscopic and imaging results in the optical and infrared. A search for the possible origin of unidentified infrared emission (UIE) features seen in the spectra is presented in Sect. 4. In Sect. 5 we present a comparison of our observations and the prediction of a three-dimensional (3D) model. A possible link between nested bipolar PNs and their binary central stars is discussed in Sect. 6. Finally, we summarise the conclusion in Sect. 7.

\begin{table*}
\caption{Summary of HST optical-infrared measurements of Hb 12}
        \centering
        \begin{tabular}{llccccc}
\hline\hline
Instrument & Filter & $\lambda_{c}$ (\AA~) & $\triangle$ $\lambda$ (\AA~) & Exposures (s) & Observation date & Program ID \\
\hline
& & \multicolumn{4}{c}{Visible} & \\
\hline
WFPC2 & F656N & 6564 & 22 & 400$\times$2 & 2007 Feb. 23 & 11093  \\
      & F656N & 6564 & 22 & 400$\times$2 & 2007 Feb. 24 & 11093  \\
\hline
& & \multicolumn{4}{c}{Near-infrared} & \\
\hline
NICMOS-3 & F175W & 17500 & 11000 & 3.63$\times$4 & 1998 Jan. 18 & 7695 \\
         & F175W & 17500 & 11000 & 3.63$\times$20 & 1998 Apr. 04 - Nov. 05 & 7903 \\
         & F175W & 17500 & 11000 & 3.63$\times$4 & 1998 Jun. 18 & 7959 \\
\hline
\end{tabular}
\label{tab1}
\end{table*}

\section{Observation and data reduction}\label{s2}

\subsection{HST optical and near-infrared imaging}

The {\it HST} observations of Hb 12 were retrieved from the Space Telescope Science Archive. The high-resolution imaging was made with optical narrow-band and NIR broad-band filters. The optical observations were observed with the Wide Field Planetary Camera 2 (WFPC2) mounted on the {\it HST} through program 11093 (PI: K. Noll) over the period from 2007 February 23 to February 24. This PN was observed with the narrow-band H$\alpha$ (F656N) filter ($\lambda_{c}$ = 6564 \AA~, $\triangle\lambda$ = 22 \AA) on the Planetary Camera (PC). The camera has a field of view (FOV) of 36$\arcsec$.8$\times$36$\arcsec$.8 with a spatial resolution of 0.$\arcsec$045 pixel$^{-1}$. Combining all the observations generates a deep image with a total integration time of 1600 s. We also retrieved the NIR broad-band images of Hb 12 obtained with the Near Infrared Camera and Multi-Object Spectrometer camera 3 (NICMOS-3) on the {\it HST} under programs 7695, 7903, and 7959 (PI: D. Calzetti). NICMOS-3 provides a FOV of 51$\arcsec$ $\times$ 51$\arcsec$ and a pixel scale of 0.$\arcsec$2 pixel$^{-1}$. The observations were performed using a broad-band F175W filter ($\lambda_{c}$ = 1.75 $\mu$m, $\triangle\lambda$ = 1.1 $\mu$m) with a total exposure time of 102 s from 1998 January 18 to November 5. We reduced and calibrated all data using standard {\it HST} pipeline procedures, including flat-field calibration and bias subtraction. Then the multiple frames of each filter were combined to improve spatial sampling and to remove cosmic rays using the STSDAS package of IRAF. A summary of the {\it HST} observations is given in Table \ref{tab1}. The processed H$\alpha$ (F656N) and NIR (F175W) images are shown in Fig.~\ref{hst}.

\subsection{CFHT near-infrared observations}

\begin{table*}
\caption{Log of CFHT WIRCam observations of Hb 12}
        \centering
        \begin{tabular}{lcccc}
\hline\hline
Filter name & $\lambda_{c}$ & $\triangle\lambda$ & Integration time & Observation date \\
            & ($\mu$m) & ($\mu$m) & (s) &  \\
\hline
H$_{2}$ S(1) $\upsilon$= 1-0 & 2.122 & 0.032 & 60$\times$4 & 2012 Aug. 4  \\
Br$\gamma$ & 2.166 & 0.03 & 57$\times$4 & 2012 Aug. 4  \\
K continuum (Kc) & 2.218 & 0.033 & 57$\times$4 & 2012 Aug. 4 \\
\hline
\end{tabular}
\label{tab2}
\end{table*}

The NIR observations were taken using the Wide-field InfraRed Camera (WIRCam) on the CFHT under RUNID 12AS98 on 2012 August 4. The WIRCam contains four 2048 $\times$ 2048 pixel HAWAII2-RG detectors and has an FOV of 20$\arcmin$ $\times$ 20$\arcmin$ with an angular resolution of 0.$\arcsec$3 pixel$^{-1}$. Three narrow-band filters, H$_{2}$ ($\lambda_{c}$ = 2.122 $\mu$m, $\triangle$ $\lambda$ = 0.032 $\mu$m), Br$\gamma$ ($\lambda_{c}$ = 2.166 $\mu$m, $\triangle$ $\lambda$ = 0.03 $\mu$m), and Kc ($\lambda_{c}$ = 2.218 $\mu$m, $\triangle$ $\lambda$ = 0.033 $\mu$m), were employed. The Kc narrow-band filter was used to subtract the continuum contribution from the images obtained with the H$_{2}$ and Br$\gamma$ filters. The total exposure times for the H$_{2}$, Br$\gamma$, and Kc observations are 240, 228, and 228 s, respectively. Our observations were performed under good sky conditions (25 percentile). The seeing conditions during the entire observation run varied between 0.$\arcsec$7 and 0.$\arcsec$5. The average full width at half maximum values of the star points in H$_{2}$, Br$\gamma$, and Kc images are 0.\arcsec55, 0.\arcsec51, and 0.\arcsec56, respectively. All data were reduced through the standard procedures in the NOAO IRAF software, including flat-field correction, basic crosstalk removal, dark subtraction, and sky subtraction using neighbouring images. The final images were produced by combining and mosaicking different exposure images with sky subtraction. The journal of the {\it CFHT} measurements is given in Table \ref{tab2}. 

\begin{table*}
\caption{Available AKARI IRC, Spitzer IRS, and ISO observations}
        \centering
        \begin{tabular}{cccc}
\hline\hline
 ID & Observation date & Wavelength range & Exposures \\
           &                  & ($\mu$m) & (s) \\
\hline
& \multicolumn{2}{c}{AKARI IRC observation} & \\
\hline
3460035 & 2009 Jul. 18 & 2.5 - 5.0 & 355   \\
\hline
& \multicolumn{2}{c}{Spitzer IRS spectra} & \\
\hline
AOR key 18183936 & 2006 Sep. 09 & 9.9 - 19.6 & 14  \\
& & 18.7 - 37.2 & 12 \\
\hline
& \multicolumn{2}{c}{ISO observation} & \\
\hline
TDT 43700330 & 1997 Jan. 26 & 2.4 - 45.2 & 1912  \\
TDT 57101028 & 1997 Jun. 09 & 43 - 195.9 & 1318  \\
\hline
\end{tabular}
\label{tab3}
\end{table*}

\begin{figure*}
\begin{center}
\includegraphics[width=0.9\textwidth]{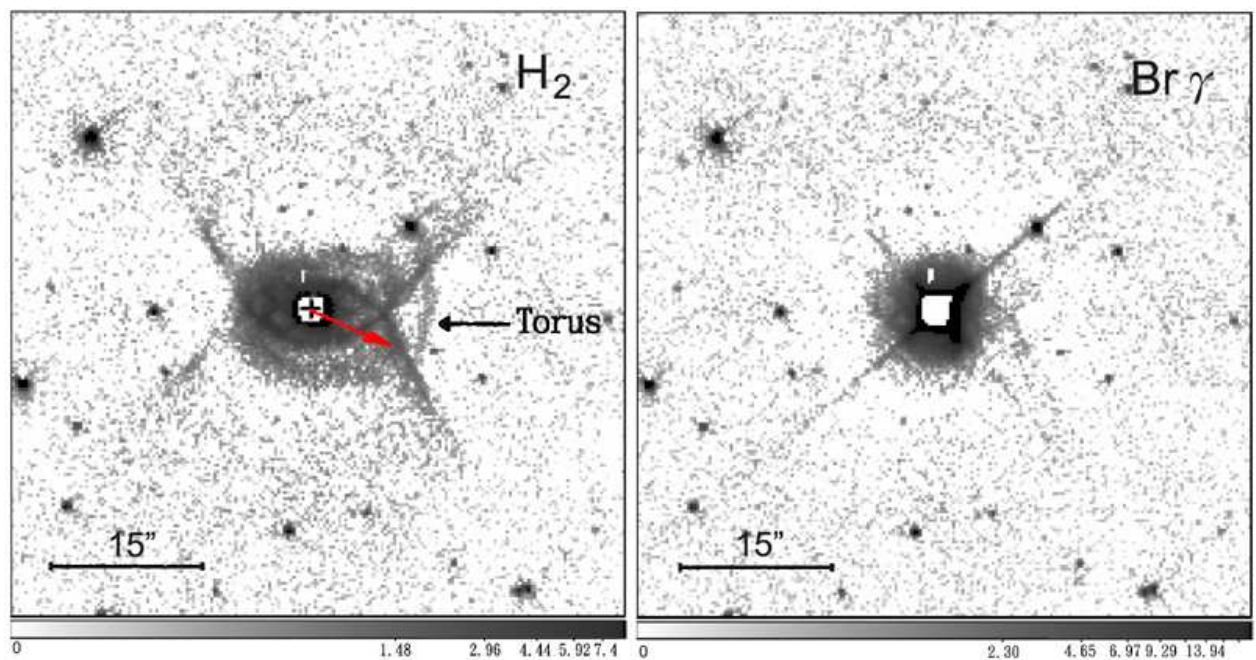}
\end{center}
\caption{Comparison between the {\it CFHT} residual H$_{2}$ and Br$\gamma$ images of Hb 12. North is up and east is to the left. The core region of this nebula is saturated. The black cross denotes the position of the central star. The proper motion direction is marked with red arrow. The intensities are in a  logarithmic scale and the scale bars shown at the bottom are given in units of 10$^{-16}$ erg cm$^{-2}$ s$^{-1}$.}
\label{cfht}
\end{figure*}

\begin{figure}
\begin{center}
\includegraphics[width=0.48\textwidth]{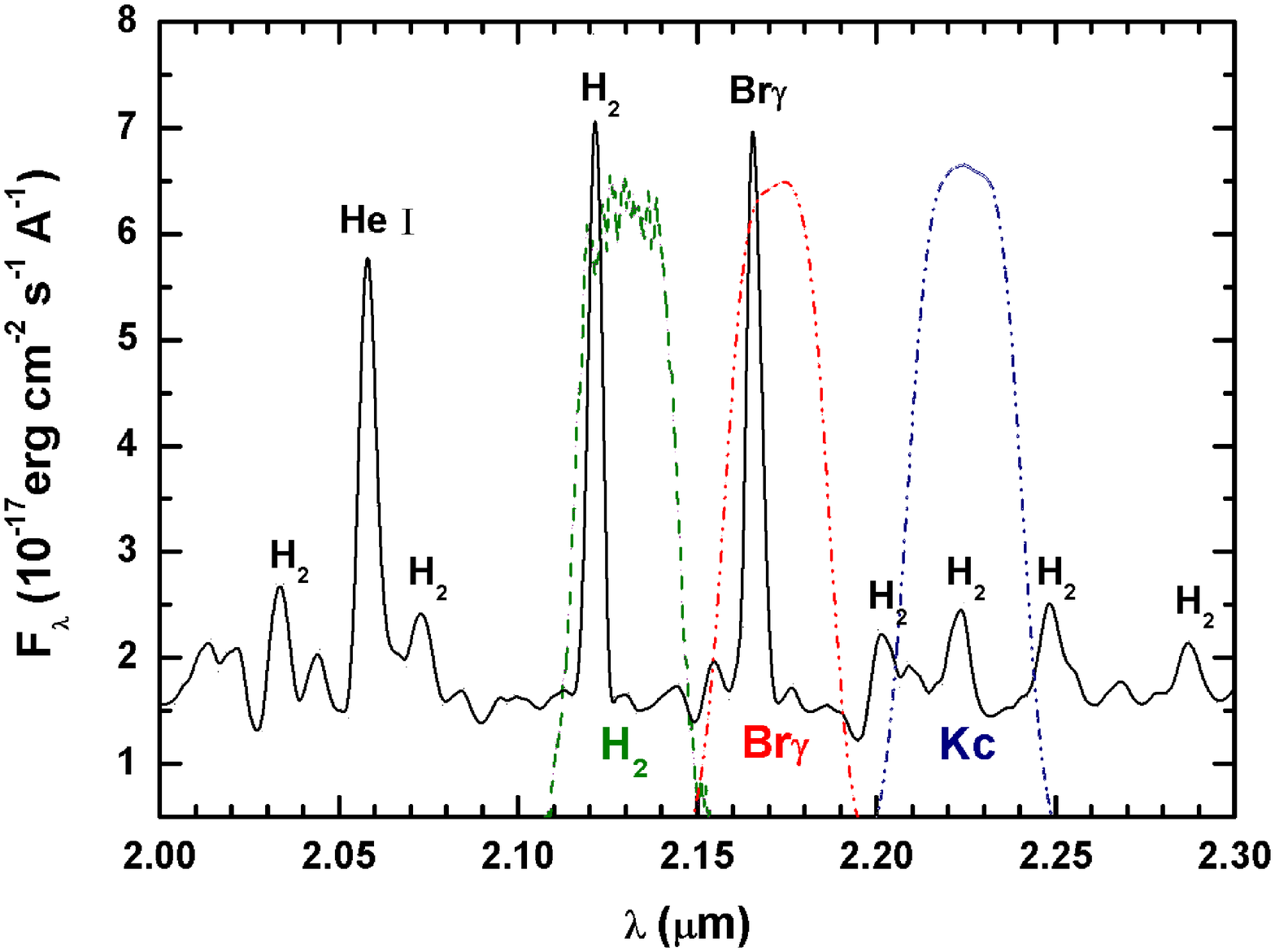}
\end{center}
\caption{Near-infrared (2.0 $-$ 2.3 $\mu$m) spectrum of Hb 12 at a position (3.\arcsec7 E, 2\arcsec S) relative to the core \citep{Hora99}. The normalised relative spectral response curves of the H$_{2}$, Br$\gamma$, and Kc filters are plotted as dashed green, red, and blue lines, respectively.}
\label{filter}
\end{figure}

While optical energy output from PNs is dominated by nebular emission lines, continuum emission could significantly contribute to the NIR narrow-band images. In order to examine the spatial distributions of molecular and ionised hydrogen of Hb 12, we subtracted the continuum contribution from the Br$\gamma$ and H$_{2}$ images using the observed Kc image scaled by factors that were obtained by comparing continuum fluxes at 2.13, 2.17, and 2.23 $\mu$m in the NIR spectrum of \citet{Hora99}. The continuum-subtracted Br$\gamma$ and H$_{2}$ images of Hb 12 are shown in Fig.~\ref{cfht}. We note that the central regions within a radius of $\sim$ 4.\arcsec0 are saturated.

Fig.~\ref{filter} shows the NIR spectrum of Hb 12 at a position (3.\arcsec7~E, 2\arcsec~S) relative to the core \citep{Hora99}. We note that the contributions from H$_2$ lines to the Kc image are not negligible. This might lead to an over-subtraction in extended regions. According to the spectral measurements, we estimate that the effects caused by the H$_{2}$ 2.223 $\mu$m are $\leq$ 15$\%$ and $\leq$ 12$\%$ for the H$_{2}$ 2.122 $\mu$m and Br$\gamma$ images, respectively. Presumably, the contamination of the H$_2$ lines has a spatial variation, but this is too small to affect our main results.

\subsection{Spitzer IRS spectra}

The mid-IR spectra of Hb 12 were obtained with the {\it Infrared Spectrograph} (IRS; Houck et al. 2004) on board the {\it Spitzer
Space Telescope} through program 30403 (PI: B. Jiang) with Astronomical Observation Request (AOR) key 18183936. The nebula was observed using the Short High (SH) and Long High (LH) modules, which cover the wavelength range of 9.9 $\mu$m - 37.2 $\mu$m with a spectral dispersion of $R\sim$ 600. The aperture sizes are 4$\farcs$7$\times$11$\farcs$3 in SH module and 11$\farcs$1$\times$22$\farcs$3 in LH module. The total on-source exposure time of the IRS observations is 26 s.

The data were reduced starting with basic calibrated data (BCD) from the {\it Spitzer} Science Centre pipeline version s18.7. We used the {\bf IRSCLEAN} program to remove rogue pixels. To extract the spectra, the Spectroscopic Modelling Analysis and Reduction Tool (SMART) analysis package \citep{Higdon04} was used. The final spectrum was made by combining the IRS observations to enhance the signal-to-noise ratio (S/N). Because the diaphragm size of SH measurement (4$\farcs$7$\times$11$\farcs$3) is smaller than the LH aperture size (11$\farcs$1$\times$22$\farcs$3), the SH slit covers a small part of the emission area while the LH diaphragm can cover almost entire nebula ($\sim$ 10$\arcsec$ in diameter; Vaytet et al. 2009), and thus a scaling correlation is needed for the SH wavelength range of IRS spectrum. By comparing the continuum flux levels at the overlapping region of the SH and LH observations, we scaled the SH measurement by a factor of $\sim$ 5.14. A journal of IRS spectroscopic measurements is summarised in Table \ref{tab3}.

\subsection{AKARI IRC and ISO spectra}

Hb 12 was observed on 2009 July 18 using the infrared camera (IRC; Onaka et al. 2007) on the {\it AKARI} satellite. The observations were processed using the observation template IRCZ4 mode in the post-liquid-helium warm phase with a spectral resolving power of {\it R} $\sim$ 100, which cover a wavelength of 2.5 - 5.0 $\mu$m. The diaphragm size is 3$\arcsec\times$ 1$\arcmin$ for the nebula with the slit-less mode, and the total exposure on source of the measurements was 355 s. The data reduction and calibration included flat fielding, linearity correction, dark and background subtractions, as well as flux and wavelength calibrations based on the standard procedure with the IRC Spectroscopy Toolkit for Phase 3 (version 20111121). The accuracies of wavelength and absolute flux calibrations were estimated to be $\sim$ 0.01 $\mu$m and $\sim$ 10$\%$, respectively \citep{Ohyama07}.

The IR spectra of this nebula were also obtained by the  {\it Infrared Space Observatory} (ISO)  through three observation programs from 1997 January 26 to 1997 June 09, for which  the Short-Wavelength Spectrometer (SWS, PI: M. Jourdain de Muizon) module (2.4 - 45.2 $\mu$m, $\lambda$/$\triangle\lambda$ = 1600) and Long-Wavelength Spectrometer (LWS, PI: M. Barlow) module (43 - 196 $\mu$m, $\lambda$/$\triangle\lambda$ = 200) were employed. The diaphragm sizes are 33$\arcsec\times$ 20$\arcsec$ and 84$\arcsec\times$ 84$\arcsec$ in the SWS and LWS modules, respectively, and thus the measurements cover the entire nebula (see Fig.~\ref{cfht}). The raw data were initiated using the {\it Offline processing software} (OLP) version 10.1. Further data reductions were performed using the ISO {\it Spectral Analysis Package} (ISAP) version 2.1 to remove questionable data points. Then the different scans for each detector were averaged to improve the S/N. 

The Spitzer IRS, AKARI IRC, and ISO SWS spectra of this object have been published by \citet{Jiang13}, \citet{Ohsawa16}, and \citet{Hora04}. Here we re-processed these IR spectra, aiming to investigate the properties of unidentified emission features and the dust components. A journal of AKARI and ISO spectroscopic observations is also included in Table \ref{tab3}.

\section{Results}\label{s3}

\subsection{Visual and near-infrared imaging of Hb 12}\label{lobes}

The {\it HST} images (Fig.~\ref{hst}) clearly reveal that this nebula has a size of $\sim$ 10$\arcsec$ and extends approximately along the north-south (N-S) direction. The bandpass of the F175W filter mainly includes the contributions from strong lines such as the \ion{H}{I} Br$\gamma$ and the H$_{2}$ 2.12 $\mu$m lines, weak lines including \ion{He}{I}, and [\ion{Fe}{II}] lines, and  weak continuum scattered by the dust. The wavelength coverage of F175W band is the widest of the {\it HST} NIR filters, thus allowing us to search for possible faint structures around this nebula. A close inspection of the F175W image reveals two distinct pairs of butterfly-shaped, open-end bipolar lobes (labelled 'ionised lobe' and 'secondary lobe'), while the inner ionised lobe dominates in the H$\alpha$ image. The F175W image reveals one more pair of lobes (marked 'primary lobe') located outside the ionised and secondary lobes, which form a complex nested eye-like features in the equatorial region. An eye-like structure like this is also visible in the {\it HST} F215N, F212N, F160W, F110W \citep{Hora00, Kwok07b, Clark14}, and H$_{2}$ images \citep{Hora96, Fang18}, which is probably due to the intersection of the inclined outer bipolar structures (primary lobe and secondary lobe). From the F175W image, we also note that the eastern side of the outermost lobe is generally fainter than the western side. This phenomenon was also revealed by the {\it HST} visual images of \citet{Vaytet09} and may indicate less dense regions on the eastern side of this PN \citep{Hyung96}. An azimuthal asymmetry can be produced during the AGB or post-AGB stages \citep{Welch99}. Alternatively, this may be due to less dust extinction on the western side of the lobe.

In addition, a faint arc filament is visible in the H$\alpha$ image (marked 'arc'), which is not visible in the F175W image. This structure might be the optical counterpart of the northern part of the largest eye-like shell. The lack of the south arc might be due to dust extinction. A series of two-dimensional (2D) rings aligned with the inner ionised lobe is also visible in the  H$\alpha$ image, as reported by \citet{Kwok07b}.

\subsection{Main structures in H$_{2}$ and Br$\gamma$ emission}\label{knots}

\begin{figure*}
\begin{center}
\includegraphics[width=1\textwidth]{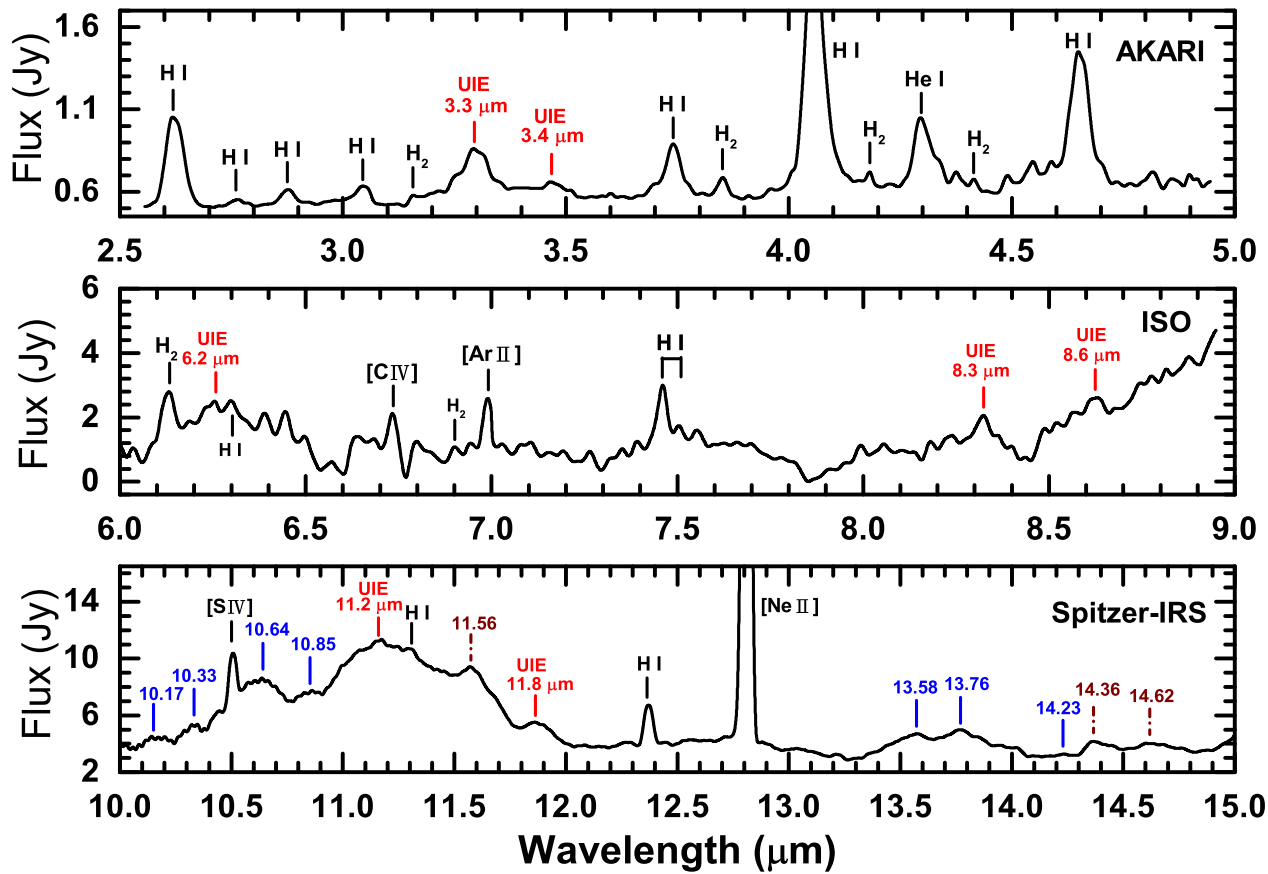}
\end{center}
\caption{Continuum-subtracted spectra of Hb 12 in wavelength ranges from 2.5 to 5 $\mu$m ({\it AKARI-IRC}, Ohsawa et al. 2016), 6 to 9 $\mu$m ({\it ISO-SWS}), and 10 to 15 $\mu$m ({\it Spitzer-IRS}). The prominent H$_{2}$ and \ion{H}{I} lines are marked. The red, blue, and brown lines mark the positions of known UIE features (3.3, 3.4, 6.2, 8.3, 8.6, 11.2, and 11.8 $\mu$m), crystalline silicate (Molster et al. 2002), and new UIE features, respectively.}
\label{iso}
\end{figure*}

The {\it CFHT} H$_{2}$ image shows that the axes of three pairs of bipolar structures have almost the same orientation. The measured position angles (PAs) of the axes of three bipolar structures are -8.5$^\circ\pm$2.3$^\circ$ \citep{Kwok07b}, -6.8$^\circ\pm$3.2$^\circ$, and -6.8$^\circ\pm$3.2$^\circ$ for the ionised, secondary, and primary lobes, respectively. The angular sizes of the ionised, secondary, and primary lobes are about 9.$\arcsec$1$\times$5.$\arcsec$6, 7.$\arcsec$1$\times$8.$\arcsec$5, and 24.$\arcsec$1$\times$32.$\arcsec$3. These lobes may originate from several episodes of mass ejection events collimated by a similar formation mechanism during the AGB or post-AGB phases.

To search for possible extended molecular and ionised structures (H$_{2}$ and Br$\gamma$ emission) around Hb 12 and understand their spatial distributions in the nebula, we generated {\it CFHT} residual images of this object to enhance the extended structures seen in H$_{2}$ and Br$\gamma$ (Fig.~\ref{cfht}). As shown Fig.~\ref{cfht}, 
the eye-shaped structures and the two outer pairs of bipolar structures (secondary and primary lobes) are dominated by H$_{2}$ line emission. The central structure extending along the N-S direction (near the core) in the residual Br$\gamma$ image is similar to the [\ion{Fe}{II}] bipolar lobe detected in the 1.64 $\mu$m image \citep{Welch99, Clark14}. The formation of these bipolar [\ion{Fe}{II}] structures is likely due to the interaction of circumstellar material and a collimated fast wind, producing shocked [\ion{Fe}{II}] emission along the side walls of the inner hourglass regions \citep{Welch99}. The distributions of atomic and molecular hydrogen are clearly spatially resolved. The inner ionised region with Br$\gamma$ emission is surrounded by two outer H$_{2}$ lobes and the eye-shaped structures.

The {\it CFHT} residual H$_{2}$ image (Fig.~\ref{cfht}, left panel) also reveals an apparent boundary near the western waist of Hb 12 (marked ``torus''), which is an indication of the interaction between the PN and surrounding ISM. The proper-motion direction of this nebula is estimated to be PA = -124$^\circ$ \citep{Urban09}, which is roughly towards the direction of the torus. Furthermore, two pairs of separate, symmetric H$_{2}$ knotty structures are found in the wide-field H$_{2}$ image and were previously reported in \citet{Hsia16} and \citet{Fang18}. These knots might be formed by ballistic ejection \citep{Palmer96} and may have originated from different separate eruptive events \citep{Vaytet09}, which may provide further evidence of a binary origin in the central star of this nebula \citep{Clark14, Hsia06, Kwok07b, Vaytet09}. These structures are frequently found around bipolar PNs and are often associated with collimated outflows with high velocity, such as Fleming 1 \citep{Lopez93, Palmer96}, NGC 6881 \citep{Guerrero98}, and MyCN 18 \citep{Bryce97, O'Connor00}. 

The angular distances between individual knots and the central source are measured to be $\sim$ 63.\arcsec2, 191.\arcsec3, 69.\arcsec7, and 161.\arcsec3 for knots $N1$, $N2$, $S1$, and $S2$ (see Fig. 2 of \citet{Hsia16}). Adopting a distance of 2.88 kpc \citep{Phillips02} for Hb 12, the deprojected physical distances from the central star to the knots are $\sim$ 0.88 pc, 2.67 pc, 0.97 pc, and 2.25 pc for knots $N1$, $N2$, $S1$, and $S2$. Assuming that the radial velocity of these knots with H$_{2}$ emission is equal to the average value of that seen in [\ion{N}{II}] ($\sim$ 121 km s$^{-1}$; Vaytet et al. 2009) and the inclination angle is 45$^\circ$  \citep{Fang18}, the derived kinematic ages of knots $N1$, $N2$, $S1$, and $S2$ are about 7140, 21600, 7880, and 18200 yr, respectively. This agrees well with an earlier suggestion that the knots were ejected at various events during the transition from AGB to post-AGB stage \citep{Vaytet09}.  

\begin{table*}
        \caption{Available photometric measurements of Hb 12}
        \centering
        \begin{tabular}{llc}
                \hline\hline
Filters & Flux & Reference \\
\hline
\multicolumn{3}{c}{Central star and nebula} \\
\hline
B (mag) & 14.5$\pm$0.5 & \citet{Tylenda91} \\
V (mag) & 13.6$\pm$0.5 & \citet{Tylenda91} \\
R$_{u}$ (mag) & 11.753 & \citet{Roser08} \\
\hline
 \multicolumn{3}{c}{Dust} \\
\hline
2MASS J (mag) & 10.234$\pm$0.02 & \citet{Cutri03} \\
2MASS H (mag) & 9.825$\pm$0.02 & \citet{Cutri03} \\
2MASS K$_{s}$ (mag) & 8.811$\pm$0.02 & \citet{Cutri03} \\
WISE F 3.4 $\mu$m$^a$ (Jy) & 0.34$\pm$0.01 & This study \\
WISE F 4.6 $\mu$m$^a$ (Jy) & 0.66$\pm$0.02 & This study \\
WISE F 12 $\mu$m$^a$ (Jy) & 16.87$\pm$0.12 & This study \\
WISE F 22 $\mu$m$^a$ (Jy) & 59.30$\pm$0.05 & This study \\
Spitzer F 3.6 $\mu$m$^b$ (Jy) & 0.38$\pm$0.01 & This study \\
Spitzer F 4.5 $\mu$m$^b$ (Jy) & 0.57$\pm$0.06 & This study \\
Spitzer F 5.8 $\mu$m$^b$ (Jy) & 0.95$\pm$0.04 & This study \\
Spitzer F 8.0 $\mu$m$^b$ (Jy) & 6.12$\pm$0.20 & This study \\
Spitzer F 24.0 $\mu$m$^b$ (Jy) & 67.66$\pm$0.35 & This study \\
MSX F 8.3 $\mu$m (Jy) & 10.99$\pm$0.45 & \citet{Egan03} \\
MSX F 12.1 $\mu$m (Jy) & 16.95$\pm$0.85 & \citet{Egan03} \\
MSX F 14.7 $\mu$m (Jy) & 18.53$\pm$1.13 & \citet{Egan03} \\
MSX F 21.3 $\mu$m (Jy) & 50.81$\pm$3.05 & \citet{Egan03} \\
IRAS F 12 $\mu$m (Jy) & 25.13$\pm$1.26 & \citet{Tajitsu98} \\
IRAS F 25 $\mu$m (Jy) & 68.91$\pm$2.76 & \citet{Tajitsu98} \\
IRAS F 60 $\mu$m (Jy) & 28.86$\pm$3.46 & \citet{Tajitsu98} \\
IRAS F 100 $\mu$m (Jy) & 11.90$\pm$1.07 & \citet{Tajitsu98} \\
AKARI F 9 $\mu$m (Jy) & 11.88$\pm$0.05 & \citet{Ishihara10} \\
AKARI F 18 $\mu$m (Jy) & 40.60$\pm$0.45 & \citet{Ishihara10} \\
AKARI F 65 $\mu$m (Jy) & 28.82$\pm$1.21 & \citet{Ishihara10} \\
AKARI F 90 $\mu$m (Jy) & 19.70$\pm$1.08 & \citet{Ishihara10} \\
AKARI F 140 $\mu$m (Jy) & 6.50$\pm$0.76 & \citet{Ishihara10} \\
AKARI F 160 $\mu$m$^c$ (Jy) & 3.08: & \citet{Ishihara10} \\
JCMT F 450 $\mu$m$^d$ (Jy) & 1.4 $>$ & \citet{Hoare92} \\
JCMT F 800 $\mu$m (Jy) & 0.311$\pm$0.048 & \citet{Hoare92} \\
JCMT F 1100 $\mu$m (Jy) & 0.305$\pm$0.028 & \citet{Hoare92} \\
\hline
 \multicolumn{3}{c}{Free-free emission} \\
\hline
90 GHz (Jy) & 0.275$\pm$0.079 & \citet{Purton82} \\
30 GHz (Jy) & 0.399$\pm$0.009 & \citet{Pazderska09} \\
22 GHz (Jy) & 0.201$\pm$0.027 & \citet{Purton82} \\
15 GHz (Jy) & 0.17 & \citet{Aaquist91} \\
10.6 GHz (Jy) & 0.122$\pm$0.02 & \citet{Purton82} \\
8.1 GHz (Jy) & 0.081$\pm$0.003 & \citet{Purton82} \\
5 GHz (Jy) & 0.0796 & \citet{Stanghellini10} \\
5 GHz (Jy) & 0.065 & \citet{Aaquist90} \\
2.7 GHz (Jy) & 0.027$\pm$0.003 & \citet{Purton82} \\
 \hline
 \multicolumn{3}{c}{IUE spectra} \\
\hline
Instrument & Wavelength range ($\mu$m) & Exposures (s) \\
\hline
SWP 05572 & 0.115 - 0.20 & 5400   \\
SWP 17075 & 0.115 - 0.20 & 9000  \\
LWR 04814 & 0.185 - 0.33 & 7200  \\
LWR 13359 & 0.185 - 0.33 & 7200  \\
\hline
 \end{tabular}
\tablefoot{
\tablefoottext{a}{Measured from the WISE all-sky data release.} \tablefoottext{b}{Measured from the Spitzer data release.} 
\tablefoottext{c}{The flux measured from AKARI at 160 $\mu$m is unreliable.} 
\tablefoottext{d}{The JCMT 450 $\mu$m flux represents an upper limit detection.}}
\label{tab4}
\end{table*}

\subsection{Chemistry of the dust component of Hb 12}\label{features}

Compared to the ISO SWS spectroscopic observation, the AKARI IRC and Spitzer SH spectra are observed with smaller apertures, which can cover parts of the nebula. These shorter-wavelength measurements need to be multiplied by scaling factors of $\sim$2.68 for the AKARI part and $\sim$7.32 for the SH slit to be consistent with the continuum flux levels at 4.8 and 12.2 $\mu$m in the ISO SWS spectrum. There is a slight flux difference between the ISO SWS and Spitzer SH spectra at about 10 $\mu$m, which is due to the larger flux uncertainties at the spectral edges. To better understand the properties of IR emission features, we fitted the continua using third-order polynomials and subtracted them from observed IR spectra. Fig.~\ref{iso} shows the {\it AKARI-IRC}, {\it ISO-SWS}, and {\it Spitzer-IRS} continuum-subtracted spectra in the 2.5 - 15 $\mu$m region of Hb 12. The most prominent bands are the fine structure lines of [\ion{C}{IV}] at 6.71 $\mu$m, [\ion{Ar}{II}] at 6.99 $\mu$m, [\ion{S}{IV}] at 10.51 $\mu$m, and [\ion{Ne}{II}] at 12.81 $\mu$m, and weaker hydrogen recombination lines of \ion{H}{I} (6-4) at 2.63 $\mu$m, \ion{H}{I} (8-5) at 3.74 $\mu$m, \ion{H}{I} (14-6) at 4.02 $\mu$m, \ion{H}{I} (7-5) at 4.65 $\mu$m, \ion{H}{I} (6-5) at 7.46 $\mu$m, and \ion{H}{I} (7-6) at 12.37 $\mu$m (Fig.~\ref{iso}). 

Interestingly, several weak UIE features at 3.3, 3.4, 6.2, 8.3, 8.6, 11.2, and 11.8 $\mu$m are visible. The 6.2, 8.3, 8.6, 11.2, and 11.8 $\mu$m bands are detected for the first time in this nebula. Also present are the silicate bands at 10.17, 10.33, 10.64, 10.85, 13.58, 13.76, and 14.23 $\mu$m \citep{Molster02}. These features reveal a mixed chemistry with oxygen and carbon dust environments, which may involve a mass-loss process affected by a binary core \citep{Waters98, Cohen99, Cohen02, Werner14} and is commonly seen in post-AGB stars \citep{Waters98, van03, de06} and young PNs \citep{Stanghellini12, Guzman15}. These results may reflect that this nebula has experienced a transition from an oxygen-rich to a carbon-rich nature.

The IR spectra were examined for possible presences of new UIE bands.
Previously, two uncommon UIE features at 14.36 and 14.62 $\mu$m were noted by \citet{Jiang13}, which are clearly shown in Fig.~\ref{iso}. We also detect an additional UIE at 11.56 $\mu$m. The characteristics of these bands are still unclear and may be related to crystalline silicates, as suggested by \citet{Jiang13}. A detailed wide laboratory and astronomical study with crystalline silicate is needed to further investigate the nature of these new UIE features.

\subsection{Spectral energy distribution}\label{s3-3}

\begin{figure}
\begin{center}
\includegraphics[width=0.48\textwidth]{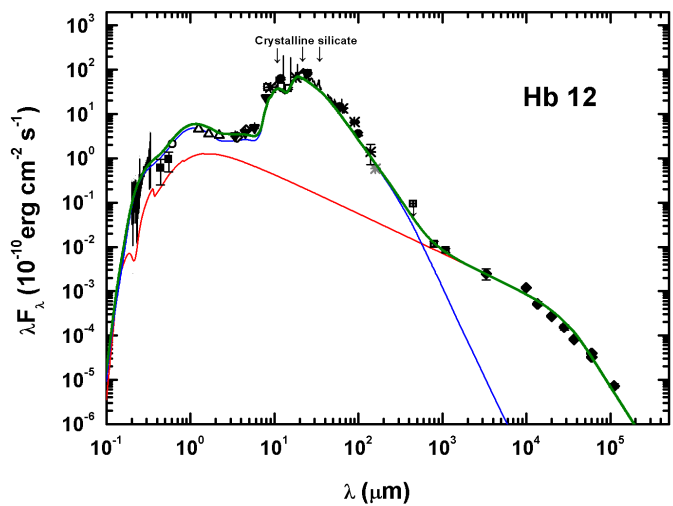}
\end{center}
\caption{SED of Hb 12 covering the wavelength range from the UV to the radio. Filled squares and open circle show the B, V, and R$_{u}$ photometric results. The open triangles, filled triangles, open diamonds, open squares, filled circles, asterisks, crossed squares, and filled diamonds show {\it 2MASS, Spitzer, WISE, MSX, IRAS, AKARI, JCMT}, and available radio measurements, respectively. The unreliable {\it AKARI} detection is shown as light grey asterisk. The flux measured from {\it JCMT} at 450 $\mu$m represents an upper limit detection. The spectra with {\it IUE, ISO LWS, \textup{and} SWS} are also plotted. Broad silicate features at 9.7, 18, 23, 28, and 33 $\mu$m are also marked. The red and blue curves represent the gaseous continuum and the emergent fluxes estimated by using a dust radiative transfer model, respectively. The sum of fluxes emitted from all components is plotted as a green curve.}
\label{sed}
\end{figure}

Strong IR excesses emitted from circumstellar dust components are often thought to be a typical characteristic of young PNs. To compare the relative contributions of dust, photospheric, and nebular components of Hb 12, the spectral energy distribution (SED) of this PN was constructed (Fig.~\ref{sed}). The visual B and V photometry of the central source and R$_{u}$ magnitude of entire nebula were obtained from 
\citet{Tylenda91} and the PPM-Extended (PPMX) database \citep{Roser08}, respectively. Near and mid-IR magnitudes are obtained from 2MASS, {\it Midcourse Space Experiment} (MSX), {\it Infrared Astronomical Satellite} (IRAS; Tajitsu \& Tamura 1998), and AKARI catalogues. Additional photometric data measured from {\it Spitzer} and {\it Wide-field Infrared Survey Explorer} (WISE; Wright et al. 2010) observations are also added using the procedure presented in \citet{Hsia14a}. In the FIR range, the fluxes obtained from the {\it James Clerk Maxwell Telescope} (JCMT) measurements were observed by \citet{Hoare92}. Also plotted in Fig.~\ref{sed} are the {\it International Ultraviolet Explorer} (IUE), {\it ISO} LWS, and SWS spectra of this nebula. A journal of the data we used is given in Table \ref{tab4}. For the broad bands, corrections should be made for the flux values given in Table \ref{tab4} and Fig.~\ref{sed}. The aperture- and colour-correction coefficients for {\it Spitzer} and {\it WISE} bands can be found in \citet{Reach05} and \citet{Wright10}.

Fig.~\ref{sed} shows that a large amount of total fluxes from this PN is mainly emitted in the IR range due to dust particles. Assuming that the dust emission is composed of a single blackbody (BB) component, the total emerging fluxes emitted from the nebula are about 9.5 $\times$ 10$^{-9}$ erg cm$^{-2}$ s$^{-1}$. Adopting the distance to Hb 12 of 2.88 kpc \citep{Phillips02}, the total luminosity is estimated to be 2,470$\pm$170 L$_{\sun}$. In addition, we find that observed dust continuum cannot be fitted well using one BB with a single temperature, and therefore the value derived above is just a lower limit.   

The total observed fluxes emitted from this PN were fitted by a model with two components composed of reddened photospheric continuum (the sum of fluxes from the central source and the nebular emission) and dust thermal emission using the same methods as presented in \citet{Hsia19}. The relative contributions of various components of Hb 12 are clearly visible in Fig.~\ref{sed}. We note that dust continuum with the far- to mid-IR range of observed SED fitting is too broad and cannot be fitted by a component with a single temperature. Therefore we used the radiative transfer code DUSTY \citep{Ivezic99} to fit the observed SED curve. The chemical composition of the dust grains was assumed to be the mixing of silicate and silicon carbide (SiC). Although these assumptions cannot fit the broad silicate features well (Fig.~\ref{sed}), they can provide an approximation to the observed dust continuum. 

Fig.~\ref{sed} shows that the excess emission flux at the WISE 5.8 $\mu$m band is higher than expected the IR continuum. This may be due to the contributions from the H$_{2}$ line and the 6.2 $\mu$m UIE feature (see Fig.~\ref{iso}). The broad crystalline silicate features at 9.7, 18, 23-28, and 33 $\mu$m are also visible in our {\it ISO} spectra.

For this fitting, our best approximations for the effective temperatures of the hot star and the cool companion are 30,000$\pm$1,600 K and  4,500$\pm$300 K, respectively. With the same distance, the total luminosity of this object is about 2,630$\pm$320 L$_{\sun}$, which is slightly higher than the value directly estimated from the SED (2,470$\pm$170 L$_{\sun}$). The lack of UV data could introduce some uncertainties.

The location of the central star in the Hertzsprung-Russell diagram indicates that Hb 12 is a young PN, which is consistent with the previous conclusions \citep{Aaquist90, Zhang90}. Furthermore, the existence of a cool companion, as shown in the SED, also agrees with previous results \citep{Corradi03a, Hsia06, Kwok07b, Clark14}.

\section{Searching for a possible candidate with UIE features}\label{s4}

To investigate the possible origin of the UIE features seen in Hb 12 (Fig.~\ref{iso}), we qualitatively discuss interpretations of the origin of each UIE band within separate spectral wavelength regions. This is done by speculating about types of bonds or the basic functional groups underlying each feature. Then some of the possible candidates of the UIE carries are suggested. Their corresponding IR emission spectra are simulated from the normal mode vibration frequencies and are compared to that of Hb 12. The preliminary examination of the UIE bands implies an organic compound origin made of carbon and hydrogen. However, other elements such as oxygen, nitrogen, and sulfur found in this object \citep{Hora96, Hyung96} may contribute to the chemical composition of the UIE carriers. The UIE features appear as a group of bands with specific wavelengths (with some variations) that are commonly observed among astronomical sources. Their origins are sought within different classes of organic compounds. These astrochemistry models are polycyclic aromatic hydrocarbon (PAH; Duley \& Williams 1981) and their derivatives hydrogenated PAH \citep{Wagner00}, PAH with side groups \citep{Kwok01}, nitrogenated PAH \citep{Hudgins04}, Oil fragments \citep{Cataldo02}, and mixed aromatic-aliphatic organic nano-particle (MAON; Kwok \& Zhang 2011). The complexity of chemical and physical conditions that led to these UIEs in astronomical resources implies that a single molecule cannot be responsible for all these features. The common practice is to search for a group of molecules within the same class of organic compounds. It has also been suggested that different molecular types can contribute to the formation of UIEs. In any of these cases, typical examples of molecular species from the mentioned class of current astrochemistry models should be tried and tested.         
\subsection{Interpretation of emission spectra}\label{interpretation}

\subsubsection{2.5 - 5.0 $\mu$m}

In the wavelength coverage of 2.5 - 5.0 $\mu$m, two UIE features at 3.3 and 3.4 $\mu$m can be detected (Fig.~\ref{iso}). From empirical and theoretical standpoints, these two bands are assigned to the stretching vibrations of aromatic/olefinic sp$^2$ and aliphatic sp$^3$ C-H bonds separately \citep{Sadjadi17}. Recently, the experimental visualisation of normal modes in a single Co(ll)-tetraphenylporphyrin (CoTPP) molecule has established the localised and specific type of vibrations for C-H bonds of prophyrin rings \citep{Lee19}. A vibration mode consisting of the stretching of three aromatic C-H bonds (two in-phase and one out-phase) on each of the phenyl rings of CoTPP was experimentally observed at 3.35 $\mu$m \citep{Lee19}. For pure hydrocarbons, these vibrations cover the wavelength range of 3.2 to 3.3 $\mu$m \citep{Sadjadi17}. A combined quantum chemical and vibrational normal mode analysis shows that the stretching vibrations of aliphatic sp$^3$ C-H bonds usually occur within the range of 3.35 - 3.47 $\mu$m for hydrocarbons without hetero-atoms (N, O, S; Sadjadi et al. 2017). Within this wavelength coverage, the analysis has successfully distinguished four different types of vibrational modes, and they are symmetric and asymmetric vibrations of C-H bonds on methyl (-CH$_{3}$) and methylene (-CH$_{2}$-) functional groups \citep{Sadjadi17}, respectively. The position of the UIE band at 3.4 $\mu$m implies that this band is more likely to originate from asymmteric vibrations of C-H bonds in methylene group \citep{Sadjadi17}.

\begin{figure}
\begin{center}
\includegraphics[width=0.48\textwidth]{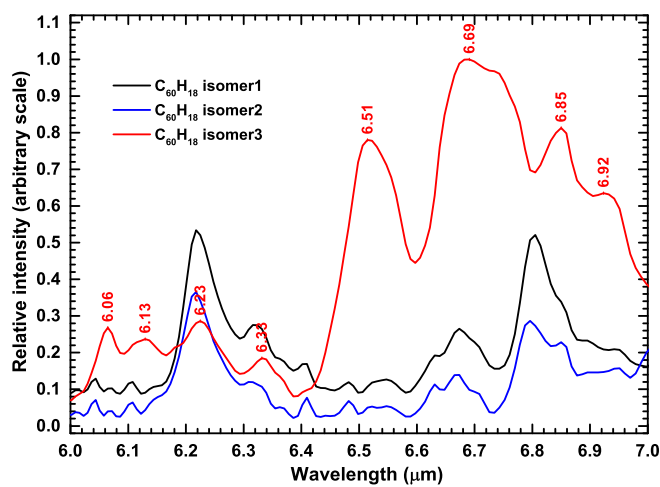}
\end{center}
\caption{Comparison of 6.0 - 7.0 $\mu$m experimental spectra of three isomers of C$_{60}$H$_{18}$ with a solid phase. Peak positions of obvious  C$_{60}$H$_{18}$ isomer3 features are marked. Spectral profiles are provided by F. Cataldo (private communication).}
\label{abdi1}
\end{figure}

\begin{figure}
\flushleft
\includegraphics[width=0.49\textwidth]{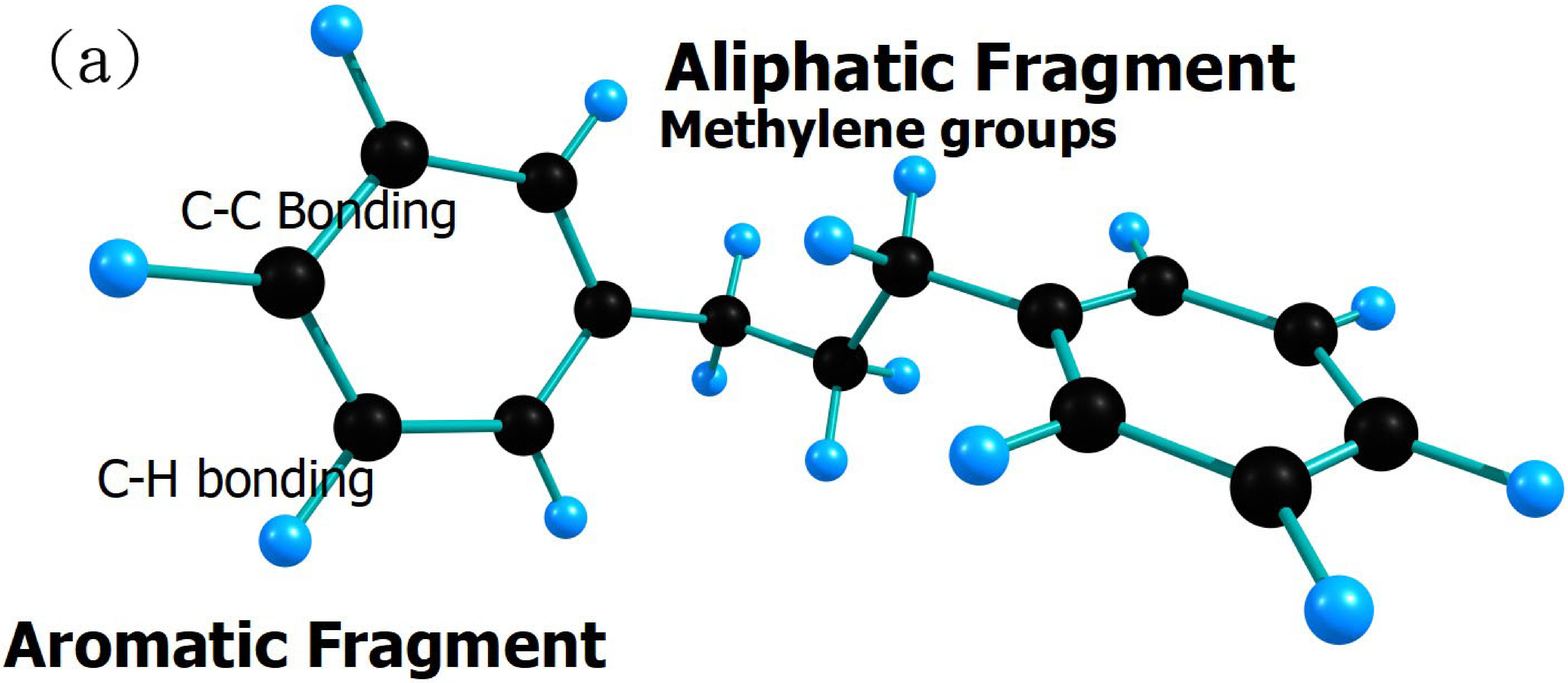}
\includegraphics[width=0.39\textwidth]{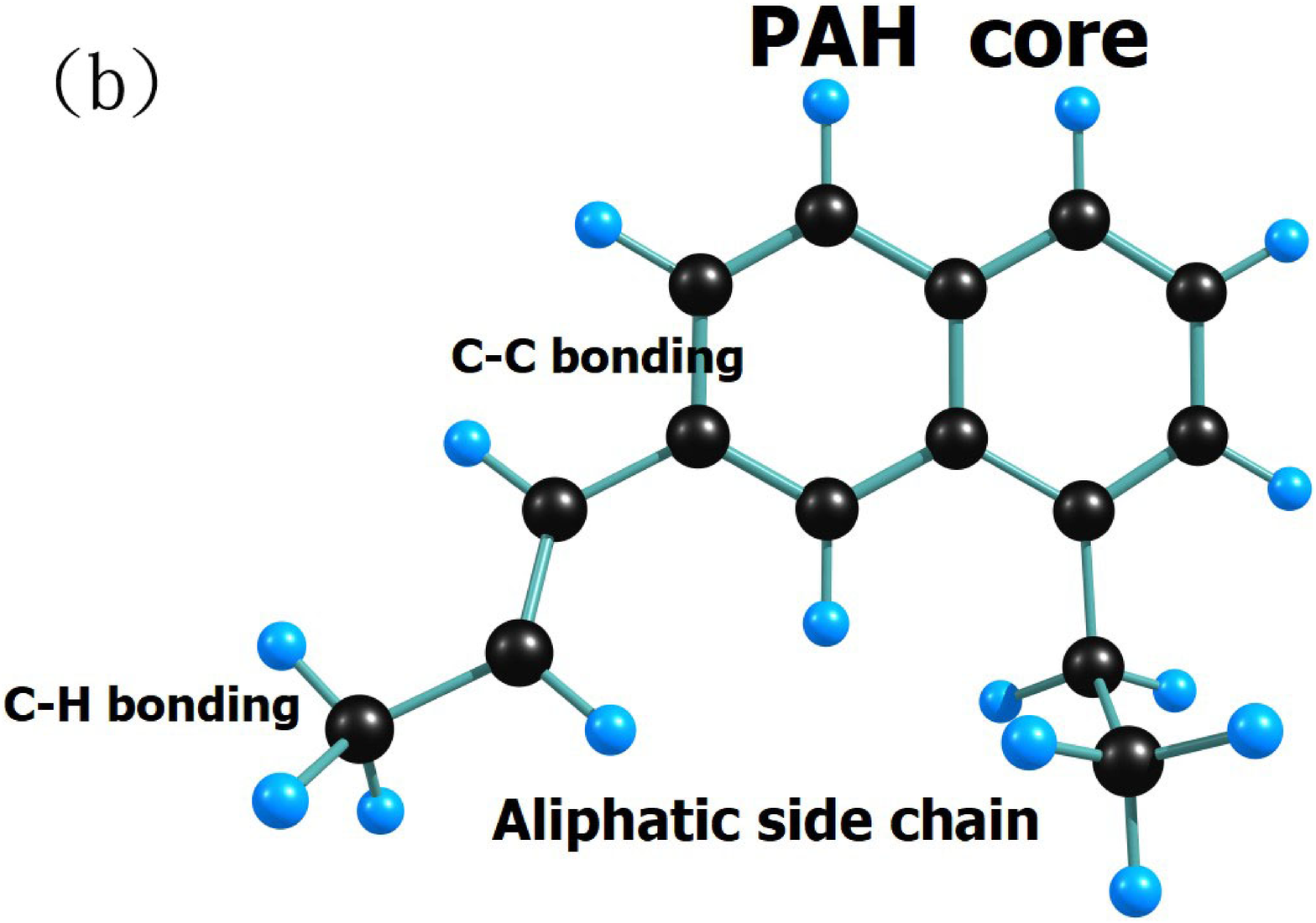}
\caption{3D structural rendering of two candidate molecules, both with the same chemical formula as C$_{15}$H$_{16}$ , composed of aromatic and aliphatic fragments. (a) MAON. (b) PAH with aliphatic side groups. The carbon and hydrogen atoms are shown in black and blue, respectively.}
\label{fig7}
\end{figure}

\subsubsection{6.0 - 9.0 $\mu$m}
 
From Fig.~\ref{iso}, three UIE features at 6.2, 8.3, and 8.6 $\mu$m are visible in the wavelength coverage (6.0 - 9.0 $\mu$m). Normal vibrational analysis demonstrated quantitatively that the appearance of 6.2 $\mu$m band is the result of the couplings between the stretching vibration of aromatic sp$^2$ C-C and in-plane bending of aromatic sp$^2$ C-H bonds \citep{Sadjadi18}. This coupling has a profound effect on the IR signature of organic carbon-based molecules. For instance, both experimental and theoretical IR spectra of C$_{60}$ fullerene lack this band, although the C-C bonds show characteristic features of aromatic bonds \citep{Zhang17}. Theoretical calculations (B3LYP/PC1 model) show that the normal mode vibrations of fullerene start from 6.4 $\mu$m to higher wavelength regions. However, laboratory experimental data show that a strong feature at 6.2 $\mu$m appears in the IR spectra upon hydrogenation of C$_{60}$ molecule (Fig.~\ref{abdi1}). Although the hybridisation of carbon should stand at sp$^3$ in C-H bonds in fulleranes, the appearance of 6.2 $\mu$m band is not consistent with it. This demonstrates the complex origin of the 6.2 $\mu$m feature and also the complex role of the two cited normal modes in creating this band in different organic molecules.

The basic bond types responsible for 8.3 and 8.6 $\mu$m are similar to that for 6.2 $\mu$m. The difference is the degree of coupling between two types of vibrations \citep{Sadjadi18}. Both bands are essentially assigned to the pure in-plane bending mode vibration of aromatic sp$^2$ C-H bonds.  

\subsubsection{10.0 - 15.0 $\mu$m}

A broad band is centred on 11.2 $\mu$m, and the following two weak bands at 11.56 and 11.8 $\mu$m are shown in Fig.~\ref{iso}. The feature around 11.2 $\mu$m is usually attributed to out-of-plane bending mode vibrations of sp$^2$ C-H bonds, which can be either part of aromatic or olefinic fragments. In the case of honeycomb PAH molecules, this type of vibration can cover a wide wavelength range, starting at 10.5 to 14.0 $\mu$m \citep{Sadjadi18}.

\begin{figure}
\begin{center}
\includegraphics[width=0.48\textwidth]{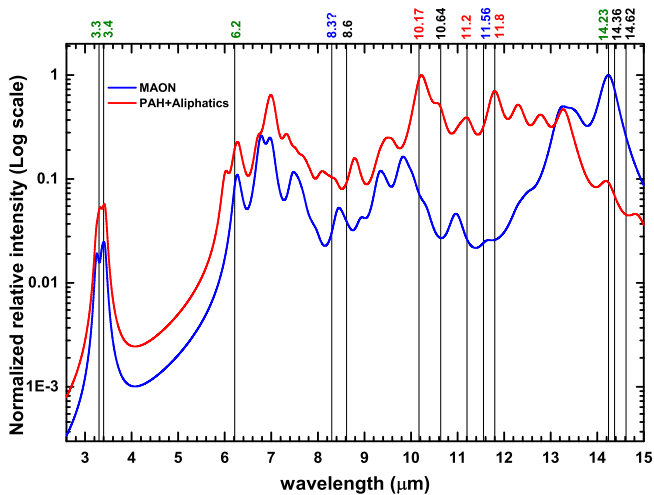}
\end{center}
\caption{Theoretically simulated IR spectra of two molecular structures with C$_{15}$H$_{16}$ formula at T = 500 K. The blue and red profiles represent MAON (1,3-diphenylpropane; Fig. 7a) and PAH with aliphatic side groups (Fig. 7b). A quantum chemical model of B3LYP/PC1 is applied to calculate the vibrational normal modes. The solid black lines indicate the positions of the UIE features seen in Hb 12. The positions of UIE features marked with green, blue, red, and black colours indicate the bands reproduced by both models, by the MAON model and the PAH+aliphatics model, and by none of the models, respectively.}  
\label{abdi3}
\end{figure}

\subsubsection{Beyond 15.0 $\mu$m}

The theoretical study of a large number of vibrational normal modes in complex hydrocarbon species (without any hetero-atoms) has provided a grand picture of the vibrational origin of UIE bands \citep{Sadjadi18}. These are bonding (6 - 14.4 $\mu$m) and skeleton (beyond 14.4 $\mu$m) origin vibrations \citep{Sadjadi18}. Although most of UIE features observed in Hb 12 such as 3.3, 3.4, 6.2, 8.3, 8.6, and 11.2 $\mu$m belong to the bonding control vibrations, the origin of two unidentified features at 11.56 and 14.36 $\mu$m is still unclear. Because these bands are located at the bonding control region of IR spectrum, they are likely due to new types of C-H bonding or hetero-atoms X with X-C or X-C-H bonds.  

\begin{figure*}
\begin{center}
\includegraphics[width=0.9\textwidth]{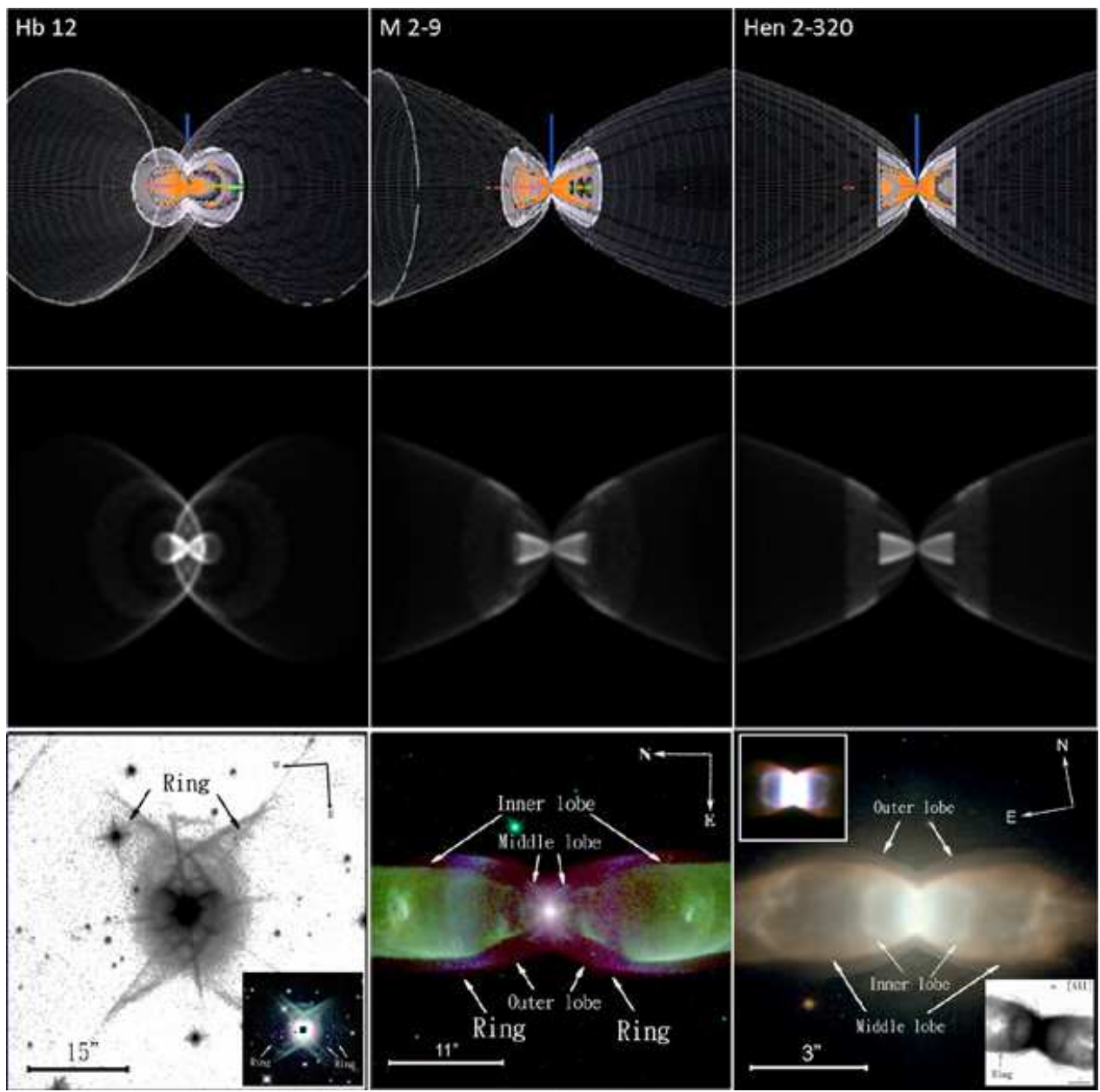}
\end{center}
\caption{Comparison of the 3D models of Hb 12 with various orientations and colour-composite {\it HST} images of PNs M 2-9 and Hen 2-320. Upper row: 3D mesh models of Hb 12 viewed at angles of 45$^\circ$, 19$^\circ$, and 0$^\circ$, respectively (0$^\circ$ being the sky plane). The innermost lobes of the models are shown in orange to better show various structures. Middle row: Corresponding rendered images of 3D mesh models. The 2D ring elements in the outermost lobes are not shown in the meshes. All renderings in this figure are applied using a Gaussian blur. Lower row, from left to right: Grey-colour {\it HST} F160W image with a composite-colour {\it CFHT} insert (made with Kc (blue), H$_{2}$ (green), and Br$\gamma$ (red) bands) of Hb 12 and composite-colour {\it HST} narrow-band images of M 2-9 and Hen 2-320 \citep{Hsia14b}.\ The {\it HST} image of M 2-9 is created with [\ion{O}{III}] (blue), [\ion{S}{II}] (green), and [\ion{N}{II}] (red) filters. The locations of the bipolar lobes and the 2D ring structures are also marked.}.
\label{model}
\end{figure*}

\subsection{A possible candidate}\label{candidate}

The simplest picture that we can anticipate for the chemical identity of the unknown carriers of the 3.3, 3.4, 6.2, 8.3, 8.6, and 11.2 $\mu$m bands seen in Hb 12 is a hydrocarbon molecule composed of both aromatic and aliphatic parts. For the origin of UIEs, a set of main UIE bands at 3.3, 6.2, 7.7, 11.3, and 12.7 $\mu$m was considered to be the characteristic features of PAH molecules \citep{Li20}. The shortcomings of the PAH model in describing different features of UIE bands have also recently been discussed in details by \citet{Jones18}. Two of UIE bands at 7.7 and 12.7 $\mu$m are not identified in the IR spectra of this nebula, indicating that Hb 12 is not typical of normal PNs with mixed chemistry (i.e. Stanghellini et al. 2012). For these reasons, we first chose a non-PAH model of astrochemistry to start our exploration of the origin of the observed UIE bands in Hb 12. This proposed model is constructed of a type of molecule candidate called MAON, which was first  hypothesised by \citet{Kwok11} as the UIE carrier in the ISM environment. It is interesting that a layer of MAON-type organic polymers was unambiguously detected inside the Solar System below the icy surface of the Saturn moon Enceladus \citep{Postberg18}. 

A simple MAON may contain two benzene rings connected by a chain of methylene (-CH$_{2}$-) groups \citep{Sadjadi16}. The 3D molecular structure and theoretically simulated IR spectrum of a proposed MAON-type candidate C$_{15}$H$_{16}$ (1,3-diphenylpropane) is shown in Figs.~7a and~\ref{abdi3}, respectively. This type of hydrocarbon molecule is composed of aromatic and aliphatic components, as shown in Fig.~7a.

The B3LYP/PC1 quantum chemical model was applied to calculate the molecular structure and IR spectra of this molecule. The calculations were performed with the Firefly 8.2.0 package\footnote{http://classic.chem.msu.su/gran/gamess/index.html} installed on the HPC2015 supercomputer platform of the University of Hong Kong. The results are visualised using the Chemcraft interface software\footnote{http://www.chemcraftprog.com}. 

To some extent, the calculated 3D molecular structure represents a good candidate of UIE carriers in the spectra of Hb 12. Although this MAON-type candidate molecule (C$_{15}$H$_{16}$) reveals IR features at around 3.3, 3.4, 6.2, 8.3 - 8.6, 11.56, and 14.23 $\mu$m, some bands between 7 to 8 $\mu$m seen in the simulated IR spectrum of this molecule are not detected in the UIE features seen in Hb 12 (Fig.~\ref{abdi3}). 

In both experimental and theoretical spectra, the IR band at 3.4 $\mu$m is the signature of the stretching vibration of aliphatic C-H bonds in form of methyl (-CH$_{3}$-) and methylene (-CH$_{2}$-) functional groups and its counterpart, that is, the feature around 6.9 -- 7.0 $\mu$m is the deformation motion of methyl or scissoring motion of methylene group. In other words, regardless of which molecular structure is chosen, these two bands always appear together. Their tight relation is like the relation between the 3.3 and 6.2 $\mu$m bands, which are assigned to the aromatic C-H and C-C stretching vibrations. 6.2 $\mu$m is due to a coupling of stretching vibrations of aromatic C-C and C-H bondings (Sadjadi \& Parker 2018). 

The mismatch between the IR signature of our proposed candidate and observational features between 7 -- 8 $\mu$m might be due to several factors: (i) the actual carrier might be slightly different in the elemental composition, molecular formula, and structure; (ii) some features might be poorly blended with atomic emission (e.g. see Fig.~\ref{iso} for the band around 7 $\mu$m, which might carry some aliphatic features, but is blended with H$_{2}$ and [\ion{Ar}{II}] lines); (iii) some particular vibrational levels are not populated enough under the physical conditions of Hb 12; and (iv) these bands in the 7 -- 8 $\mu$m wavelength range might be partly absorbed by other molecular species or dust particles. 

In order to consider a PAH-based model (with two key UIE bands at 7.7 and 12.7 $\mu$m lacking), we designed and calculated the IR spectrum of another structural isomer of the MAON C$_{15}$H$_{16}$ molecule. This is a derivative of naphthalene molecule with side aliphatic branches (Fig.~7b), and the simulated IR spectrum of this type molecule is also shown in Fig.~\ref{abdi3}. Our main purpose is to take these molecules as examples to test whether the observed UIE features are reproduced by both models. A comparison between the simulated spectra of MAON- and PAH-based models (Fig.~\ref{abdi3}) shows that although the predicted positions of UIE features in these two models are not exactly the same as the bands observed in Hb 12, some UIE features such as the 3.3, 3.4, 6.2, and 14.23 $\mu$m bands are successfully reproduced by both models, while 10.17, 11.2, and 11.8 $\mu$m features can be seen in the simulated spectrum of PAH-based model. In terms of relative intensities and peak positions, the theoretical bands in both models both agree and disagree with the observation spectra (Fig.~\ref{abdi3}). For instance, the 3.4 $\mu$m bands appear slightly stronger than the 3.3 $\mu$m band, and the bands around 7 $\mu$m can be also seen in both models. This agrees with laboratory experimental data, but not with what is observed in the spectra of this nebula. The intensity ratios of the 3.3 to 3.4 $\mu$m bands may be affected by the molecular structures, the numbers of aromatic, aliphatic, and olefinc C-H bondings, and the local physical conditions of these molecules \citep{Sadjadi17}. Based on all these factors together, our conclusion is that the UIE carrier or carriers might inherit some spectroscopic features of these two models, although spectroscopic features are not by any means the additive properties.

\begin{table*}
        \caption{Comparison of measured and simulated parameters of the observed features}
        \centering
        \begin{tabular}{lccccccc}
                \hline\hline
      & \multicolumn{3}{c}{Observed} & & \multicolumn{3}{c}{Model} \\
       \cline{2-4}\cline{6-8}
Feature & P.A.$^a$ & Size & Inclination && P.A.$^a$ & Absolute dimensions & Inclination \\
        & ($^\circ$) & ($l\arcsec$$\times$$w\arcsec$) & ($^\circ$) && ($^\circ$) & ($l\arcsec$$\times$$w\arcsec$) & ($^\circ$) \\
\hline
Ionized lobe & -8.5$\pm$2.3$^b$ & 9.1$\times$5.6$^b$ & 38$^b$ && -7 & 9.0$\times$5.5 & 45 \\
Secondary lobe & -6.8$\pm$3.2$^c$ & 7.1$\times$8.5$^c$ & ... && -7 & 7.0$\times$8.3 & 45 \\
Primary lobe & -6.8$\pm$3.2$^c$ & 24.1$\times$32.3$^c$ & 45$\pm$7$^d$ && -7 & 24.1$\times$32.3 & 45 \\ 
\hline
\end{tabular}
\tablefoot{
\tablefoottext{a}{Measured from the major axis orientation.}
\tablefoottext{b}{Measured from the {\it HST} [\ion{N}{II}] image, see Kwok \& Hsia (2007).} \\
\tablefoottext{c}{Measured from the {\it CFHT} H$_{2}$ image.}
\tablefoottext{d}{From Fang et al. (2018).}}
\label{tab5}
\end{table*}
 
\section{Simulated 3D structures of Hb 12}\label{s5}

The source Hb 12 is a young PN with nested shells along almost the same direction. Other members belonging to this family include Hen 2-104, M 2-9 \citep{Kwok07b}, and Hen 2-320 \citep{Hsia14b}. The nested bipolar lobes of these objects may arise because the consecutive ejections are collimated by the same mechanism \citep{Hsia14b}. To understand the geometry of the intrinsic structures observed in the optical-infrared images, we performed a synthetic model using the \textit{SHAPE} software \citep{steffen11}, which is a practical tool that is used to construct the 3D structures of the studied nebulae and compare the 2D appearances projected onto the sky with observations. We performed no hydrodynamic or radiative transfer simulation because our purpose is to study the complex 3D geometric structures of this PN. The PAs and dimensions of bipolar structures are given according to the qualitative analysis of our {\it HST} and {\it CFHT} observations. The integrated emission intensity ($\int n_e^2 d\ell$) along the line of sight was used to represent the surface brightness. We did not consider any excitation process in our model.

The model of Hb 12 consists of three pairs of bipolar lobes with openings at the tips, which represent three observed shells (ionised lobe, secondary lobe, and primary lobe) seen in the F175W image (Fig.~\ref{hst}). The lobes of this nebula appear to have a shell-shaped structure, thus the geometry of a thin shell (0.\arcsec1) was adopted for all bipolar structures. Three pairs of lobes are assumed to share a common tilt axis, and their inclination angle was set to be 45$^\circ$ (the sky-plane inclination is 0$^\circ$; Fang et al. 2018). Furthermore, we constructed 2D ring-like structures with a uniform density distribution, which were set to lay in the outermost lobe. The centres of these structures were arranged on the axis of the largest bipolar lobe. The model parameters of the observed features are given in Table \ref{tab5}, and a comparison between the images and the corresponding 3D representations is shown in the left panels of Fig.~\ref{model}. In all rendered images, the Gaussian blurring is used to account for the observation beam. Our modelling reproduces the most obvious features (e.g. the equatorial eye-like structures and 2D-ring components) in the images well.

\begin{table*}
        \caption{Properties of nested bipolar nebulae}
        \centering
        \begin{tabular}{llcc}
                \hline\hline
PN G & Name & Binarity & Outer knots \\
                \hline
010.8+18.0 & M 2-9 & Yes$^{a, b}$ & -- \\
061.3+03.6 & M 1-91 & Yes$^c$ & -- \\
068.1+11.0 & ETHOS 1 & Yes$^d$ & Yes$^d$ \\
086.5-08.8 & Hu 1-2 & Possible$^e$ & Yes$^e$  \\
086.9-03.4 & Ou 5 & Yes$^f$ & -- \\
111.8-02.8 & Hb 12 & Yes$^g$ & Yes$^h$ \\
300.7-02.0 & Hen 2-86 & -- & -- \\
307.5-04.9 & MyCn 18 & Yes$^i$ & Yes$^j$ \\
315.4+09.4 & Hen 2-104 & Yes$^{k, l}$ & Yes$^{k, l}$ \\
331.7-01.0 & Mz 3 & Yes$^m$ & -- \\
352.9-07.5 & Hen 2-320 & -- & -- \\
-- & M 2-56 & -- & Yes$^n$ \\
        \hline
        \end{tabular}
        \tablefoot{
        \tablefoottext{a}{Lykou et al. (2011).} 
        \tablefoottext{b}{Castro-Carrizo et al. (2012).} 
        \tablefoottext{c}{Rodr\'iguez et al. (2001).} 
        \tablefoottext{d}{Miszalski et al. (2011).}
        \tablefoottext{e}{Fang et al. (2015).} \\
        \tablefoottext{f}{Corradi et al. (2014).} 
        \tablefoottext{g}{Hsia et al. (2006).}
        \tablefoottext{h}{Fang et al. (2018).}
        \tablefoottext{i}{Miszalski et al. (2018).}
        \tablefoottext{j}{Sahai et al. (1999).} 
        \tablefoottext{k}{Corradi et al. (1993).}  \\
        \tablefoottext{l}{Santander-Garc\'ia et al. (2008).} 
        \tablefoottext{m}{Akras et al. (2019).}
        \tablefoottext{n}{S\'anchez Contreras et al. (2010).}
        }
\label{tab6}
\end{table*}

The results suggest that the eye-like structure can be produced by the intersection of tilted bipolar structures without the need of an equatorial torus. The model can predict the nebular appearance when viewed from different perspectives, as shown in Fig.~\ref{model}. For comparison, we show the observed images of another two nested PNs, M 2-9 and Hen 2-320, which can be reproduced by the 3D model with an inclination angle of 19$^\circ$ and 0$^\circ$, respectively. The inclination angles agree earlier results \citep{Hsia14b, Schwarz97}. A comparison between the renderings and the observed images of M 2-9 and Hen 2-320 (Fig.~\ref{model}) shows that although the aspect ratios and shapes of these lobes in the two PNs are not exactly the same as in the model of Hb 12, some basic structures such as the 2D ring components and shell-like features can be successfully reproduced by our model. When the modelling is tilted at angles of 19$^\circ$ and 0$^\circ$, three distinct pairs of bipolar lobes are clearly visible, and the modelled images reveal almost edge-on 2D ring features laid in the lobes. The appearances of these rotated models are similar to the apparent structures found in M 2-9 and Hen 2-320, although some input parameters needed to be adjusted to provide an exact match. Therefore we infer that these nested PNs may have similar intrinsic structures and a common origin. Further kinematic measurements are required to validate this, for which high-resolution spectroscopy is called for.

\section{Link between nested bipolar nebula and the binarity of the central star}\label{s7}

The mechanism responsible for the morphology transitions during the
AGB-PN evolution is a long-standing open problem. Interactions in binary nuclei might play a major role in the shaping of PNs. The nested hourglass PNs belong to a particular group of young bipolar nebulae, which are in a short transitional stage evolving towards developed bipolar lobe \citep{Clark14}. The nested bipolar structures could be (i) produced by multiple eruptive mass ejection events in the common-envelope (CE) evolutionary stage \citep{Corradi14}, (ii) the result of practically coeval lobes, which is similar to the PN Hen 2-104 \citep{Corradi01}, or (iii) the result of an ionisation gradient excited by far-UV photons from the central source, as suggested in PN M 2-9 \citep{Smith05}.

The appearance of highly structured bipolar lobes and two pairs of H$_{2}$ knots of Hb 12 \citep{Hsia16, Fang18} suggests that multiple pulsed mass-loss episodes have occurred during the AGB or post-AGB phases \citep{Clark14, Vaytet09, Welch99}. It would be interesting to investigate whether the nested bipolar structures of PNs are correlated with the binarity of their central stars. For this purpose, we collected the reports of 12 PNs that are known to show nested bipolar nebulae  \citep{Akras19, Castro12, Corradi93, Corradi14, Fang15, Hsia06, Lykou11, Misz11, Misz18, Rodriguez01, Sanchez10, Sahai99, Sant08}, as listed in Table \ref{tab6}. The PNG numbers and target names are given in Cols. 1 and 2. Columns 3 and 4 show whether these PNs have binary nuclei and reveal outer knots.

Table \ref{tab6} shows that a significant fraction ($\sim$ 67$\%$) of these hourglass nebulae have binary nuclei, and 6 out of 12 objects (ETHOS 1, Hb 12, Hen 2-104, Hu 1-2, M 2-56, and MyCn 18) reveal outer knotty structures. The binary central stars and outer knots detected in these PNs (Table \ref{tab6}) supports the hypothesis that multiple ejection events with time-dependent collimated winds from a binary system can form nested nebular structures
\citep{Balick19, Dennis09}.

\section{Summary}

The {\it HST} images of Hb 12 show three pairs of bipolar lobes and a faint H$\alpha$ arc filament, which appears to be the optical counterpart of the northern part of the eye-like structure. Diffuse nebulosities near the western waist of Hb 12 are revealed in the residual $H_{2}$ image. We detected the filaments that resulted from the interaction between this PN and the surrounding ISM. The IR spectra of this PN exhibit weak UIE bands at 3.3, 3.4, 6.2, 8.3, 8.6, 11.2, and 11.8 $\mu$m and silicate features at 10.17, 10.33, 10.64, 10.85, 13.58, 13.76, and 14.23 $\mu$m. These features show a mixed chemistry with both oxygen and carbon dust environments, which may involve the mass-loss process affected by a binary system. We discussed a MAON-like molecule,  C$_{15}$H$_{16}$, as the possible carrier of UIE. Although this molecule alone cannot account for the UIE phenomenon, its IR spectrum shares some similarity with the observed spectrum. From the analysis of the SED from UV to radio channels, we discovered a cool companion with a temperature of 4,500 K in the central star of this nebula. These results can provide further observational evidence for the existence of a binary system in the nucleus of Hb 12 \citep{Hsia06}. To understand the 3D internal structures of Hb 12, a model enabling the visualisation of this PN at various orientations was constructed. The modelling results show that Hb 12 may resemble other nested hourglass nebulae, Hen 2-320 and M 2-9, suggesting that this type of PNs may be common and that the morphologies of PNs are not so diverse as the morphology shown by their 2D appearances. Furthermore, we found that 8 out of the 12 nested bipolar nebulae have binary nuclei. It is worth to investigate the association between the bipolar nebulae with nested shells and the spectral properties of their central stars in the future. The appearance of the nested shell and the phenomenon of a mixed chemistry in Hb 12 leads us to suggest that this PN results from interacting nested winds that emanate from the central binary.

\begin{acknowledgements}

We are grateful to the referee Dr. William Henney for the comments that helped us to greatly improve this paper. This research uses data obtained through the Telescope Access Program (TAP), which is funded by the National Astronomical Observatories, Chinese Academy of Sciences, and the Special Fund for Astronomy from the Ministry of Finance. Part of the data presented in this paper were obtained from the Multi-mission Archive at the Space Telescope Science Institute (MAST). STScI is operated by the Association of Universities for Research in Astronomy, Inc., under NASA contract NAS5-26555. Support for MAST for non-HST data is provided by the NASA Office of Space Science via grant NAG5-7584 and by other grants and contracts. The computations were performed using research computing facilities offered by Information Technology Services, University of Hong Kong and the Laboratory for Space Research was established by a grant from the University Development Fund of the University of Hong Kong. Financial support for this work is supported by the grants from The Science and Technology Development Fund, Macau SAR (file no: 0007/2019/A). Y. Zhang thanks the finanical supports from National Science Foundation of China (NSFC, Grant No. 11973099) and the science research grants from the China Manned Space Project (NO. CMS-CSST-2021-A09 and CMS-CSST-2021-A10). S. Sadjadi acknowledges the supports from the opening fund of State Key Laboratory of Lunar and Planetary Sciences, Macau University of Science and Technology (Macau FDCT grant No. 119/2017/A3). H. J. Han was supported by The Science and Technology Development Fund, Macau SAR (file no: 0073/2019/A2).

\end{acknowledgements}

%\clearpage

%\bibliographystyle{aa}
%\bibliography{Hb12.final}

\end{document}